\documentclass[superscriptaddress,amsmath,amssymb,aps,prl,twocolumn,reprint,floatfix]{revtex4}
\pdfoutput=1
\usepackage{amssymb,amsmath,bbm,braket,indentfirst,bm}
\usepackage{dcolumn}
\usepackage{color}
\usepackage{bigstrut}
\usepackage[pdftex]{graphicx}
\usepackage{graphicx}
\usepackage{framed}
\usepackage{setspace}
\usepackage{lineno}
\usepackage[colorlinks,linkcolor={blue},citecolor={blue},urlcolor={blue}]{hyperref}
\usepackage{color,soul}

\begin{document}

\title{Controlling energy landscapes with correlations between minima}
\author{Sai~Teja~Pusuluri}
\affiliation{Department of Physics and Astronomy and Nanoscale and Quantum Phenomena Institute, Ohio University, Athens, OH, 45701, USA}

\author{Alex~H.~Lang}
\affiliation{Physics Department, Boston University, Boston, Massachusetts 02215, USA}
\affiliation{Center for Regenerative Medicine, Boston University, Boston, MA, 02215}

\author{Pankaj~Mehta}
\affiliation{Physics Department, Boston University, Boston, Massachusetts 02215, USA}
\affiliation{Center for Regenerative Medicine, Boston University, Boston, MA, 02215}

\author{Horacio~E.~Castillo}
\affiliation{Department of Physics and Astronomy and Nanoscale and Quantum Phenomena Institute, Ohio University, Athens, OH, 45701, USA}
\affiliation{Corresponding author: castillh@ohio.edu}
\date{\today}
\begin{abstract} 

Neural network models have been used to construct energy landscapes for modeling biological phenomena, in which the minima of the landscape correspond to memory patterns stored by the network. Here, we show that dynamic properties of those landscapes, such as the sizes of the basins of attraction and the density of stable and metastable states, depend strongly on the correlations between the memory patterns and can be altered by introducing hierarchical structures. Our findings suggest dynamic features of energy landscapes can be controlled by choosing the correlations between patterns.\end{abstract}

\maketitle




Neural network models based on spin glass physics~\cite{Mezard, Toulouse1986, Bray1980, To1985} such as the Hopfield
model~\cite{Hopfield1982,Amit1985,H.Gutfreund*1988, Sourlas2007} and the Kanter and Sompolinsky (KS)
model~\cite{Kanter1987, Amit1987a}, provide a mathematical framework 
to generate a complex energy landscape using a given set of
patterns as its global minima
. Many processes, especially
in biology, can be mapped to dynamical behavior in such landscapes. For
example, cellular differentiation can be pictured in terms of
Waddington's landscape~\cite{Waddington1957} and the evolution of
species can be visualized in terms of a fitness landscape~\cite{Sew,Wright1930,Wright,Wright1982}.
Neural network models have been used to model complex
processes such as protein folding~\cite{Bryngelson1995, Bryngelson1987, Onuchic2004,Frauenfelder1988,Onuchic1997,Nakagawa2006}, biological
evolution~\cite{Kaufman}, cancer evolution~\cite{Szedlak2014, Li2016,Maetschke2014,Huang2009,Cheng2012,Huang2011} and HIV virus evolution~\cite{Ferguson2013}. 
In~\cite{Pusuluri2015a}, we used an
epigenetic landscape model~\cite{Lang2014} based on the KS Hamiltonian
to study cellular interconversion dynamics, and we were able to identify
key aspects of it {such as a one-dimensional reaction coordinate during cellular reprogramming}.

In many of the biological applications of neural-network style landscape
models, the patterns corresponding to the attractors are often highly
correlated to each other, with a complex, often hierarchical, correlation
structure~\cite{Rammal1986, Mezard1985}. The general features of these landscapes,
such as the basin sizes and the density of states, have a direct
relation with the dynamics of the corresponding biological
systems. While there has been considerable work on the energetics of these models \cite{Amit1985,amit1992modeling, hertz1991introduction} 
(including for hierarchical random memories \cite{feigelman1987augmented}), as far as we are aware, 
there are no studies in the literature about how to control dynamical features for this kind of models. 
In this letter, we study how to control those features for neural network models with
correlated minima, focusing in particular on choices of sets of
patterns that are relevant for modeling cellular interconversion
dynamics~\cite{Lang2014,Pusuluri2015a}. In our study, we emphasize how
those general features vary with the strength and structure of the
correlations, and we show that the landscape features can
be controlled by making choices among different possible hierarchical
correlation structures.


We represent each state of the system by a spin vector $\vec{S} =
(S_i)_{i=1,\cdots,N}$, where the index $i$ labels individual Ising spins
$S_i = \pm 1$, and $N$ is the total number of spins in the system. We
construct the energy landscape by choosing $p$ states $\vec{\xi}^{\mu}$
with $\mu=1,\cdots,p$ to be attractor patterns. Clearly, each component
$i$ of a given pattern $\mu$ is an Ising spin, $\xi_i^{\mu} = \pm 1$. To
quantify the proximity between patterns, we use the correlation matrix
$A^{\mu\nu} \equiv \vec{\xi}^{\mu} \cdot \vec{\xi}^{\nu}$, where the
overlap (or dot product, or correlation) of two arbitrary state vectors
is defined by $\vec{S} \cdot \vec{S}' \equiv \frac{1}{N} \sum_{i=1}^N
S_i S'_i$. 

Although the correlation matrix $A^{\mu\nu}$ gives a complete
description of the correlation between patterns, it is useful to also
define an average overlap of a given pattern $\vec{\xi}^{\mu}$ with all
others, given by $\sigma^{\mu} \equiv \frac{1}{p-1} \sum^{p}_{\nu=1, \nu
  \neq \mu} A^{\mu\nu}$. We also define the overlap $m^{\mu} \equiv
\vec{S} \cdot \vec{\xi}^{\mu}$ between a state vector $\vec{S}$ and a
given pattern $\vec{\xi}^{\mu}$. Since the patterns that we will
consider will in some cases be highly correlated to each other, it will
often be the case that if a state vector $\vec{S}$ approaches a certain
pattern $\vec{\xi}^{\mu}$, it will have a large overlap not just with
that pattern but also with many others. This makes the overlap $m^{\mu}$
a poor measure of proximity between state vectors and patterns, and it
motivates the introduction of the projection $a^{\mu} \equiv
\sum^{p}_{\nu = 1}(A^{-1})^{\mu\nu} m^{\nu}$. Unlike the overlap, the
projection of each pattern on any other pattern is exactly zero. This is
useful because when a vector $\vec{S}$ approaches a certain pattern
$\vec{\xi}^{\mu}$, its projection on that pattern converges to 1,
and its projections on all other patterns converge towards 0.

To construct each energy landscape, we use a spin-glass inspired neural
network model, with a hamiltonian $H = - \frac{1}{2}\sum_{i,j=1}^N
J_{ij} S_i S_j$, written in terms of a connection matrix 
$J_{ij}$. This matrix needs to be defined in such a way that the chosen
patterns $\vec{\xi}^{\mu}$ are indeed attractors. In the Hopfield (H)
model, $J_{ij} = J^{H}_{ij} \equiv \frac{1}{N} \sum_{\mu=1}^p
\xi_i^{\mu} \xi_j^{\mu}$. In the Kanter-Sompolinsky (KS) model, $J_{ij}
= J^{KS}_{ij} \equiv \frac{1}{N} \sum_{\mu=1}^p \sum_{\nu=1}^p
\xi_i^{\mu} \left(A^{-1} \right)^{\mu\nu} \xi_j^{\nu}$, where $A^{-1}$
is the inverse of the overlap matrix $A$. In both cases, the energy of
the attractor patterns used to generate the connection matrix is the
same for all of the patterns, and they are the only ground states in the
system. For an arbitrary state $\vec{S}$, its energy can be written in
term of its overlaps $m^{\mu}$ and projections $a^{\mu}$. In the
Hopfield case, $H^{(H)}= -\frac{N}{2} \sum_{\mu=1}^p m^{\mu}
m^{\mu}$. In the Kanter-Sompolinsky case, $H^{(KS)}= -\frac{N}{2}
\sum_{\mu=1}^p m^{\mu} a^{\mu}$.

To study the properties of the energy landscapes, we perform Monte Carlo
(MC) simulations on both the H and the KS models, using the heat bath
algorithm 
with random asynchronous
updates. The total number of independent MC runs is $Q=20000$. For each
run, we initially set the temperature at $T=\infty$, and then we
instantly quench the system by setting $T=0$. Thus, the state of the
system at the end of each quench is an attractor of the $T=0$ dynamics,
and it is either a metastable state (i.e., stable against flipping one
spin, but with higher energy than the ground states) or a stable state
(i.e. a ground state, one of the patterns used to create the
landscape). 

From the outcome of these simulations, we extract two independent pieces
of information, both of which are important for the dynamics in the
given landscape: the fraction $R_l$ of MC runs that ends in the $l$-th
attractor (which we use as a working definition of the size of this
attractor's basin of attraction), and the number $\omega(E)$ of
(metastable or stable) attractors in an energy interval from $E$ to
$E+\Delta$.
%
The definition of $\omega(E)$ is analogous to the definition of the
microcanonical partition function $\Omega(E)$. The difference between
$\omega(E)$ and $\Omega(E)$ is that the former counts only metastable
and stable states, while the latter counts {\em all\/} states. From
$\omega(E)$, we define the {\em configurational entropy\/} $S(E) \equiv
\log(\omega(E))$. 
%
To define $R_l$, we start by counting the number $Q_l$ of runs that end
in a particular attractor, labeled by an integer $l$, with $l =
1,\cdots,p$ corresponding to the $p$ stable states (the original
patterns), and $l=p+1,\cdots,p+p_m$ corresponding to the $p_m$
metastable states of the $T=0$ dynamics. With these definitions, $Q =
\sum_{l=1}^{p+p_m} Q_l$. The (relative) size of the basin of attraction
corresponding to attractor $l$ is defined by $R_l = Q_l/Q$. We can
separately combine the sizes of the basins of attraction for stable
states $R^{\text{(st)}} \equiv \sum_{l=1}^{p} R_l$, and the sizes of
the basins of attraction for metastable states, $R^{\text{ms}}
\equiv \sum_{l=p+1}^{p+p_{m}} R_l$; such that $R^{\text{(st)}} +
R^{\text{ms}}=1$. 


We begin to study the effect of correlations by comparing the landscapes
generated by two sets of patterns, one in which the patterns are highly
correlated and the other in which they are completely independent. The
first landscape was used in~\cite{Pusuluri2015a} to study the dynamics
of cell reprogramming. It is generated by a set of $p=63$ patterns (the
{\em biological patterns\/}), where each pattern $\mu$ corresponds to
one mouse cell type, and each one of the $N=1436$ spins $\xi_i^{\mu}$ in
pattern $\mu$ represents the (binarized) expression level for one gene
for that particular cell type, with $\xi_i^{\mu} = +1$ indicating that
gene $i$ is expressed, and $\xi_i^{\mu} = -1$ indicating that gene $i$
is not expressed~\cite{Pusuluri2015a}. The second is a set of $p=63$
patterns (the {\em independent patterns\/}) where each component
$\xi_{i}^{\mu}$ for $i=1,\cdots,N=1436$ and $\mu = 1,\cdots,p$ is an
independent random variable that has equal probability to have the
values $+1$ and $-1$. 

\begin{figure}[h!]
\centering
\includegraphics[width=0.5\textwidth]{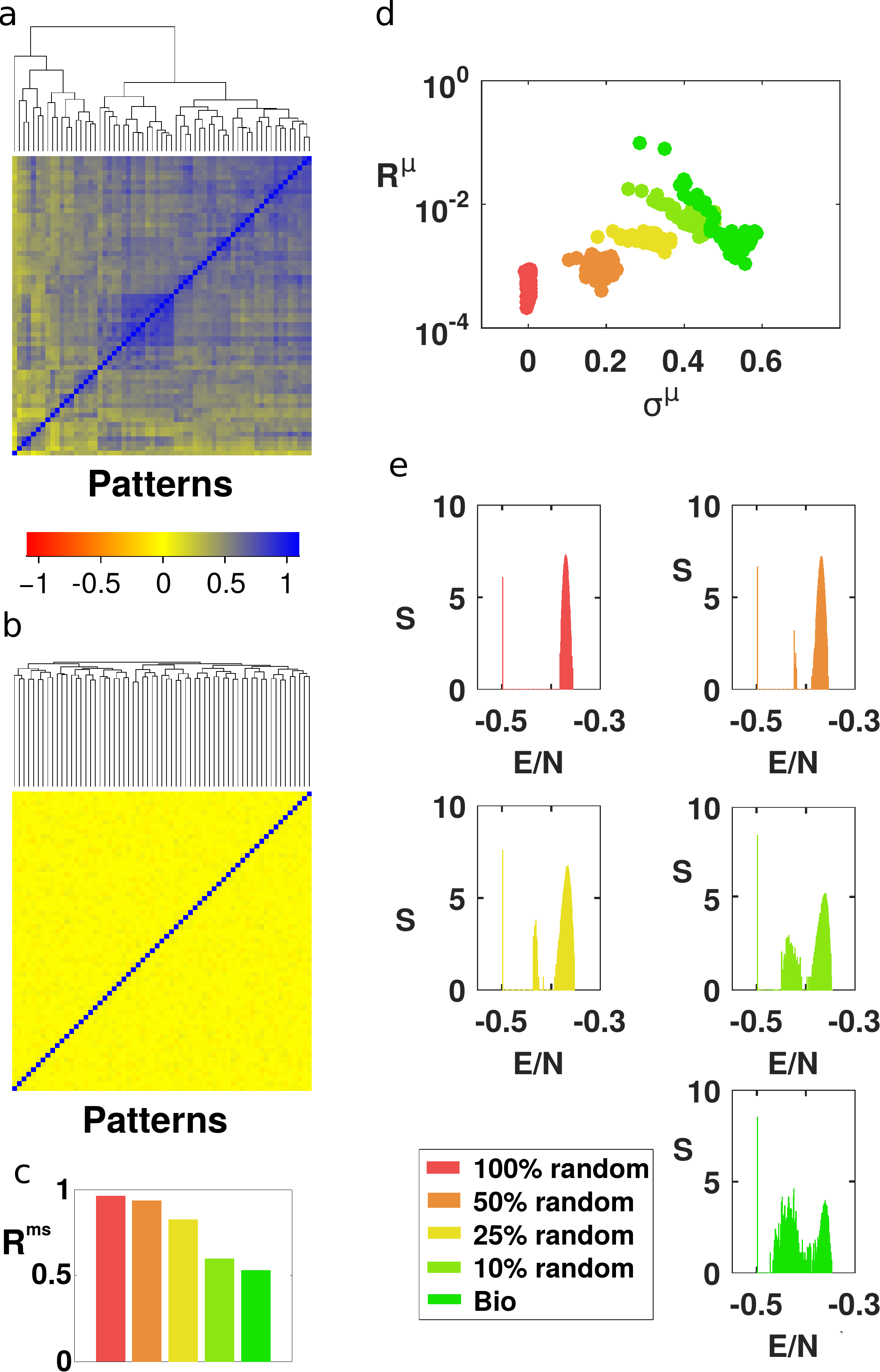}
\caption{(Color online) \textbf{Independent patterns and biological patterns.} a)
  Dendrogram and correlation matrix for the biological patterns. b)
  Dendrogram and correlation matrix for independent random
  patterns. c) Bar plot of sum of the basin sizes of all the metastable states $R^{\text{ms}}$ d) Scatter plot of the basin size $R^{\mu}$ as a function
  of the average overlap $\sigma^{\mu}$. e) Configurational entropy
  $S$ as a function of the energy per spin $E/N$.} 
\label{fig-2}
\end{figure}

Panels a and b in Fig.~\ref{fig-2} show the dendrograms and the
correlation matrices for the biological and independent patterns
respectively, color coded so that blue represents $A_{\mu \nu} = 1$ and
yellow represents $A_{\mu \nu} = 0$.
In the biological case, the overlaps between different
patterns are predominantly positive, with a median of $0.51$, while in
the case of independent patterns, the distribution of overlaps is
centered around $0$, with a standard deviation of order $N^{-1/2}$. 
The dendrogram in Fig.~\ref{fig-2}a demonstrates the existence of a
tree-like hierarchical correlation structure between the biological
patterns. The cell types that correspond to initial stages of
development are clustered together and tend to have low $\sigma^{\mu}$
values. The cell types that correspond to later stages of development
are also clustered together and tend to have high $\sigma^{\mu}$ values.
By contrast, the dendrogram in Fig.~\ref{fig-2}b shows that the
correlation structure is trivial, except for subdominant effects due to
the ${\cal O}(N^{-1/2})$ correlations between the independent patterns.

Three additional sets of patterns are generated by interpolation:
starting from the biological patterns, some fraction of the spin values
is replaced by independent random values. We consider the cases in which
10\%, 25\% and 50\% of the spins are replaced; if 100\% of the spins
were replaced we would recover the case of independent patterns. The
effect of replacing the original spin values by independent randomly
chosen ones is twofold: the average overlap between patterns is reduced,
and the correlation structure is also modified.
Once the five pattern sets have been selected, we generate the landscapes in
each one of the cases by using the KS model. 

Panels d and e in Fig.~\ref{fig-2} show the results from the simulations
in the biological case, the independent case, and the three
interpolations between them. Fig.~\ref{fig-2}d is a scatter plot of the
basin size $R^{\mu}$ as a function of the average overlap
$\sigma^{\mu}$, for all stable attractors $\mu =1,\cdots,p$, in the five
cases. The spread in the values of $R^{\mu}$ is largest for the
biological case, it initially gets much smaller as random spins are
added to the patterns, it is minimum for the 25\% randomness case, and
then it shows a slight increase as independent random spins continue to
be added, up to 100\%. We also observe that the typical size of a basin
of attraction is maximal in the biological case and it decreases with
the increase in the randomness fraction. For the biological case, there
is a strong anticorrelation between the basin size $R^{\mu}$ and the
average overlap $\sigma^{\mu}$. For the 10\% random case there is a
similar - although slightly weaker - anticorrelation, but for all the
other cases, there is at most a weak correlation, or no correlation at
all. In the biological case, the patterns with low $\sigma^{\mu}$, which
correspond to the cell types characteristic of the initial stages of
development, have bigger basin sizes compared to all other patterns. Fig.~\ref{fig-2}c shows the total size
$R^{\text{ms}}$ of the metastable basins of attraction, for
each of the five cases. We find that $R^{\text{ms}}$ is lowest
for the biological patterns, and it gradually increases as randomness is
added, up to a maximum value of $0.96$ in the case of independent random
patterns.  
Unlike in the case of stable states, we find that, for all metastable
states ($l=p+1 \ldots p+p_{m}$) , in all of the landscapes discussed in this work, $Q_l \leq 1$,
and therefore $R_l \alt 5 \times 10^{-5}$.

Fig.~\ref{fig-2}e shows plots of the contribution to the configurational
entropy $S$ from the metastable states as a function of the energy per
spin $E/N$, for all five cases. A trivial contribution $S_{\text ground}
= \log(p)$ due to the $p$ ground states at $E/N = -0.5$ is not included.
In each of those plots there is a narrow peak starting at the ground
state energy $E/N = -0.5$ from metastable states with energies barely
above the ground state energy. The rest of the features in the plots
correspond to higher-energy metastable states. In the case of
independent random patterns, the only other feature is an
inverted-parabola peak centered at $E/N = -0.375$, corresponding to a
gaussian peak in 
$\omega(E)$. This second peak is also present in the Hopfield model, and
there it is associated with the presence of spurious states constructed
from 3 ground states by majority rule. In the case of 50\% random spins,
the only changes from the case of independent random patterns is that
the second peak becomes slightly wider, and a third peak appears. This
third peak is centered at $E/N \approx -0.423 N$ 
and is very narrow. Gradually, as the fraction of
random spins is reduced, both the second and the third peaks become
wider, until eventually, for the biological case, they merge into a
single region in the energy axis, albeit still with maxima near the
locations of the second and third peak for $50\%$ randomness.

\begin{figure}[h!]
\centering
\includegraphics[width=0.5\textwidth]{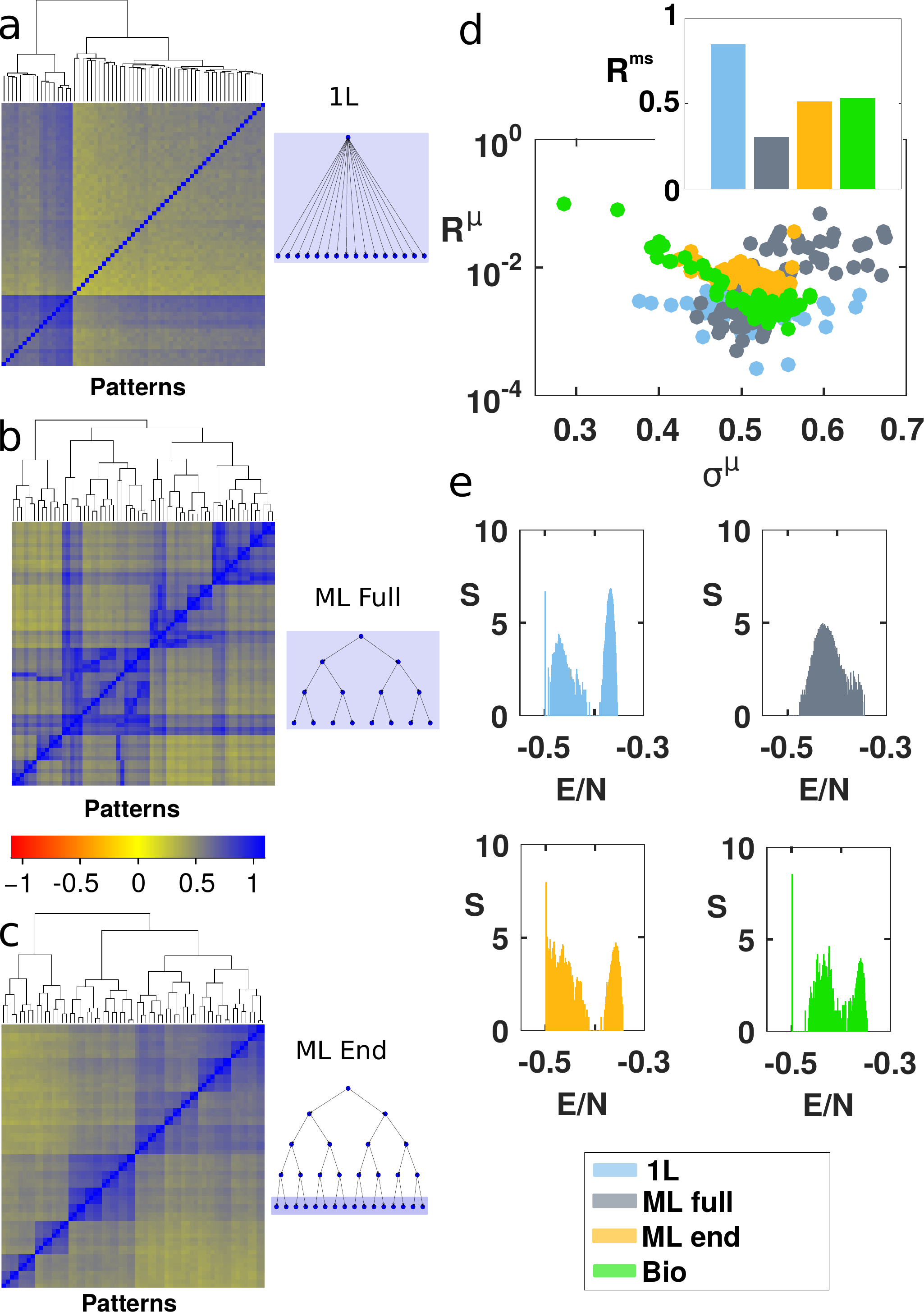} 
 \caption{(Color online) \textbf{Correlation structure}  a) Dendrogram,
correlation matrix and pattern generation schematic for the {\em 1L\/} patterns. b) Dendrogram,
correlation matrix and pattern generation schematic for the {\em ML Full\/} patterns. c) Dendrogram,
correlation matrix and pattern generation schematic for the {\em ML End\/} patterns. d) Scatter plot of the
basin size $R^{\mu}$ as a function of the average overlap
$\sigma^{\mu}$. The inset shows the corresponding $R^{\text{ms}}$ value e) Configurational entropy $S$ as a function of the energy per spin $E/N$}
  \label{fig-3}
\end{figure}

From these results it is clear that simultaneously changing the average
correlation between patterns and the correlation structure produces
substantial changes in both the configurational entropy and in the
relationship between basin sizes $R^{\mu}$ and average overlaps
$\sigma^{\mu}$. In order to separate the effects of changing the average
correlation from the effects of changing the correlation structure, we
discuss in what follows some examples in which the average correlation
between patterns is kept constant, but the correlation structure is
varied. We consider three cases. In each of the cases, we generate a
tree of patterns, starting from a randomly generated root. The first
case, labeled {\em 1L\/} (Fig.~\ref{fig-3}a), corresponds to a one-level
tree created by deriving multiple daughter patterns directly from the
root. The second case, labeled {\em ML Full\/} (Fig.~\ref{fig-3}b), is a
multi-level hierarchical tree of correlated patterns, inspired by
biological cell differentiation, where each pattern in all of the tree
levels but the last one spawns two daughter patterns, and all the nodes
in the tree are included in the pattern set. The third case, labeled
{\em ML End\/} (Fig.~\ref{fig-3}c), is a similar multi-level hierarchical
tree, but in this case only the patterns in the last level in the tree
are included in the pattern set. 
 In all three cases each daughter
pattern is generated from its parent by randomly picking a subset of the
spins and randomly generating new values for those spins. The size of
the subset of spins that is re-generated is chosen from a gaussian
distribution. The mean and average of this gaussian distribution is
adjusted in each of the three cases so that the distribution of overlaps
between the resulting patterns has the same median and standard
deviation as the distribution of overlaps between patterns in the
previously discussed biological case. Figs. \ref{fig-3}a, \ref{fig-3}b,
and \ref{fig-3}c respectively show the overlap matrices for the {\em 1L\/}, {\em ML
Full\/}, and {\em ML End\/} cases.  Once the patterns are selected, the KS model is
used to construct the landscapes for the three cases.

Panels d and e in Fig.~\ref{fig-3} show the results from the simulations
in these three cases plus the biological case. Fig.~\ref{fig-3}d shows
that in the {\em ML End\/} case, as in the biological case, $R^{\mu}$
and $\sigma^{\mu}$ are negatively correlated, while in the {\em ML
  Full\/} case, they are positively correlated, and in the {\em 1L\/}
case there is no clear correlation. The inset of Fig.~\ref{fig-3}d shows
that $R^{\text{ms}}$ has very similar values for the {\em ML
  End\/} and biological cases, but it is significantly higher for the
{\em 1L\/} case and significantly lower for the {\em ML Full\/}
case. Fig.~\ref{fig-3}e shows plots of the  metastable state 
configurational entropy 
as a function of the energy per spin $E/N$. In all cases, the
metastable states span most of the energies from just above the ground
state energy at $E/N = -0.5$ up to $E/N \approx -0.35$, with the
exception of some relatively narrow gaps. In the {\em 1L\/} and {\em
  ML End\/} cases, the gap is located around $E/N \approx -0.4$, while
in the biological and {\em ML Full\/} cases, the gap starts either at
or slightly above the ground state energy and spans up to $E/N \approx
-0.48$ or $E/N \approx -0.46$. In the {\em ML Full\/} case there is
just one maximum for $E/N \approx -0.43$. In the biological case, the
$S(E)$ curve has a two-peak shape, which is similar to the {\em 1L\/}
and {\em ML End\/} cases. For example, in all three cases there is a
peak (a remnant of the one for independent patterns) at $E/N \approx
-0.37$, separated from a continuum of states at lower energies by a
gap in the {\em 1L\/} and {\em ML End\/} cases and by a deep minimum
in the biological case. Overall, we learn from these results that the
correlation structure has a substantial effect both on the density of
states and on the basin sizes, and that out of the random tree
structures we proposed, {\em ML End\/} is the one that produces the
most similar results to the biological case. Despite the fact that in
the {\em ML Full\/} and {\em ML End\/} cases the patterns come from
the same kind of hierarchical tree, the fact that they are extracted
from different levels of the hierarchy is enough to produce dramatic
differences between the landscape features in the two cases.

Energy landscape features often play a crucial role in modeling
complex phenomena, including biological phenomena. We have found that
the general features of the landscape, such as the density of states
and the basin sizes, can be controlled by tuning the correlation
strength and by changing the correlation structure between minima.
The range of basin sizes for stable minima is strongly influence both
by the correlation strength and by the correlation structure of the
patterns.  The shape of the density of states $\omega(E) = \exp(S(E))$
and the relationship between basin size $R^{\mu}$ and average
correlation $\sigma_{\mu}$ are extremely sensitive to the correlation
structure between patterns: they are very different for
biological patterns, 
one-level trees, and for each kind of 
multi-level tree. In fact, both properties change dramatically by
simply switching from drawing patterns from all levels of a randomly
generated multi-level tree to drawing them only from the last level of
a similarly generated tree. In particular, by drawing the patterns
from the last level of a randomly generated tree, it is possible to
construct a landscape that has strikingly similar properties to the
landscape generated by using biological patterns as the minima.
We believe that these insights help lay the foundation for controlling
the properties of energy landscapes, and thus they are a first step on
the way to designing landscape models with desired dynamical
properties~\cite{Pusuluri-dynamics}. \\


\begin{acknowledgments}
\noindent {\bf Acknowledgements:} We thank members of the Boston University Center for Regenerative
Medicine (CReM) for extremely useful discussions. STP acknowledges the
Condensed Matter and Surface Science (CMSS) program at Ohio University
for support through a studentship. AHL was supported by a National
Science Foundation Graduate Research Fellowship (NSF GRFP) under Grant
No. DGE-1247312. PM was supported by the Simon's Investigator in the
Mathematical Modeling in Living Systems program and NIH grant R35 GM119461. This work was supported in part
by Ohio University.
\end{acknowledgments}

\bibliography{PRL_new2}

\newpage

\end{document}